# Comment on 'Improving the security of protocols of quantum key agreement solely using Bell states and Bell measurement'

**Jun Gu and Tzonelih Hwang**[*]

*Department of Computer Science and Information Engineering, National Cheng Kung University, No. 1, University Rd., Tainan City, 70101, Taiwan, R.O.C.*

[*]**Corresponding Author:**

Tzonelih Hwang

Distinguished Professor

Department of Computer Science and Information Engineering,

National Cheng Kung University,

No. 1, University Rd.,

Tainan City, 70101, Taiwan, R.O.C.

Email: hwangtl@ismail.csie.ncku.edu.tw

TEL: +886-6-2757575 ext. 62524

# Abstract

Designing a quantum key agreement (QKA) protocol is always a challenging task, because both the security and the fairness properties have to be considered simultaneously. Recently, Zhu et al. (Quantum Inf Process 14(11): 4245-4254) pointed out that Shukla et al.'s QKA protocol (Quantum Inf Process 13(11): 2391-2405) has some security flaws (which lead to the Participant Attack). Moreover, they proposed an improvement to avoid these weaknesses. However this study points out that the improved protocol also suffers from a colluding attack, i.e., two dishonest participants in the protocol can collaborate to manipulate the final secret key without being detected.

# 1 Introduction

In 1984, Bennett and Brassard proposed the first protocol of quantum key distribution (QKD) [1]. After that, several QKD protocols have been proposed [1-9]. QKD protocol helps the participants to establish a secret key through quantum channels. However the shared secret key is determined by the sender or the third party and then distributed to the participants. In a QKD protocol the participants cannot contribute equally to the final key. Hence in 2004 Zhou et al. proposed the first quantum key agreement protocol [10]. In contrast to the QKD protocols, each participant in QKA protocols can equally contribute their private key to establish the final secret key. The balance of the participants' contribution to the final key is thus a very serious consideration in the design of QKA protocols. And the design of a fair QKA has become an imperative research topic in quantum cryptography.

In 2010, Tsai et al. [11] pointed out that Zhou et al.'s protocol [10] cannot achieve fairness property. That is, one participant in the protocol has the ability to manipulate the final secret key without being detected by the other participant. In the same year, a QKA protocol based on the BB84 was proposed by Chong et al. [12]. In 2013, Shi et al. presented a multi-party quantum key agreement (MQKA) protocol [13] based on the Bell state and Bell measurement. Subsequently, Liu et al. [14] pointed out Shi et al.'s protocol [13] is not secure and instead proposed another MQKA protocol with single particles. In 2013, Sun et al. [15] proposed an improvement to improve the efficiency of Liu et al.'s protocol [14]. However, Huang et al. in [16] pointed out Sun et al.'s improvement cannot achieve fairness property.

In 2014, Shukla et al. proposed two QKA protocols based on Bell state and Bell measurement [17]. The first one is a two-party QKA protocol and the other is an MQKA protocol. However, recently, Zhu et al. [18] pointed out that Shukla et al.'s protocol suffers from several weaknesses and subsequently they proposed an improved three-party QKA protocol, which, they claimed, can ensure both the fairness and an enhanced level of security. However, this article shows that Zhu et al.'s protocol cannot achieve the fairness property.

The rest of this paper is organized as follows. Section 2 provides a brief review of Zhu et al.'s protocol. Section 3 analyzes Zhu et al.'s protocol and points out that two dishonest participants in the protocol can collaborate to manipulate the final secret key. Finally, a brief conclusion is given in Section 4.

## 2 Brief review of Zhu et al.'s protocol [18]

In this section, we briefly review Zhu et al.'s QKA protocol, in which three participants Alice, Bob and Charlie are involved using four different kinds of Bell state

$$|\Phi^+\rangle = \frac{1}{\sqrt{2}}(|00\rangle + |11\rangle), |\Phi^-\rangle = \frac{1}{\sqrt{2}}(|00\rangle - |11\rangle), |\Psi^+\rangle = \frac{1}{\sqrt{2}}(|01\rangle + |10\rangle), |\Psi^-\rangle = \frac{1}{\sqrt{2}}(|01\rangle - |10\rangle)$$

to establish a secure and fair key. This protocol consists of the following steps:

**Step 1** Alice prepares $n$ pairs of $|\Phi^+\rangle$ and divides them into two sequences $p_A$ and $q_A$. $p_A$ and $q_A$ are composed of all the first particles and the second particles of $|\Phi^+\rangle$ pairs. Alice also prepares two random binary bit sequences $K_A = \{0,1\}_A^{\otimes n}$ and $R_A = \{0,1\}_A^{\otimes n}$. $K_A$ can be considered as Alice's key. Similarly, Bob and Charlie generate $\{p_B, q_B, K_B, R_B\}$ and $\{p_C, q_C, K_C, R_C\}$.

**Step 2** Alice generates $\frac{n}{2}$ pairs of $|\Phi^+\rangle$ as decoy photons and concatenates them with $q_A$ to form a new sequence $q_A'$. Subsequently, Alice applies a permutation operation $\prod_A^{2n}$ on $q_A'$ to get a new sequence $q_A''$ and sends $q_A''$ to Bob. Similarly, Bob generates $q_B''$ and sends it to Charlie, Charlie generates $q_C''$ and sends it to Alice.

**Step 3** After receiving the authentic acknowledgment of receipt from Bob through the authenticated channel, Alice announces the details of permutation operation $\prod_A^{2n}$. Bob picks the decoy photons out and applies Bell measurement on them. If the error rate is found to be within the tolerable limit, they continue to the next step, otherwise they abort the protocol. At the same time, Bob and Charlie check the transmission, Charlie and Alice check the transmission.

**Step 4** After having discarded all decoy photons, Bob obtains $q_A$ according to the details of permutation operation $\prod_A^{2n}$. Subsequently, Bob performs $I$ or $X$ on $q_A$ according to $K_B$. Then Bob applies another additional unitary operation $I$ or $X$ according to $R_B$ to get a new sequence $q_{Ab}$. i.e. Bob performs $I$ or $X$ on $q_A$ according to $K_B \oplus R_B$. After this, Bob concatenates $q_{Ab}$ with decoy photons and applies permutation operation $\prod_B^{2n'}$ on them to obtain a new sequence $q_{Ab}''$.

Subsequently, Bob sends $q''_{Ab}$ to Charlie. Similarly, Charlie generates $q''_{Bc}$ and sends it to Alice, Alice generates $q''_{Ca}$ and sends to Bob.

**Step 5** The same as **Step 3**, participants check whether there is any eavesdropper exists in the transmissions

**Step 6** Follow the similar way as **Step 4**, Charlie obtains $q_{Ab}$ and performs $I$ or $X$ on it to get a new sequence $q_{Abc}$. After this, Charlie concatenates $q_{Abc}$ with decoy photons and applies permutation operation $\prod_C^{2n''}$ on them to obtain a new sequence $q''_{Abc}$. Subsequently, Charlie sends $q''_{Abc}$ to Alice. Similarly, Alice sends $q''_{Bca}$ to Bob, Bob sends $q''_{Cab}$ to Charlie.

**Step 7** The same as **Step 3**. participants check whether there is any eavesdropper exists in the transmissions.

**Step 8** Bob and Charlie first announce the details of the additional unitary operation $R_B$ and $R_C$. Charlie announces the details of the permutation operation $\prod_C^{2n''}$ after having known $R_A$ and $R_B$. Subsequently, Alice discards all decoy photons and rearranges the received sequence to obtain $q_{Abc}$. Then Alice performs Bell measurement on $\{p_A q_{Abc}\}$ to get the values of $M_A = K_B \oplus R_B \oplus K_C \oplus R_C$. Obviously, Alice can obtain the value of $K_B \oplus K_C$ according to $\{M_A, R_B, R_C\}$. Then Alice generates the final shared secret key $K = K_A \oplus K_B \oplus K_C$. Similarly, Bob and Charlie can obtain the final key $K$.

## 3 Problem with Zhu et al.'s QKA

According to Zhu et al. [18], the fairness property of a QKA protocol specifies that all involved participants can equally influence the final shared secret key. i.e. no non-trivial subset of the participants can manipulate the final shared secret key.

In this section, we try to show that Zhu et al.'s protocol cannot achieve the fairness property. That is, the final shared secret key can be manipulated by two dishonest participants. Let us assume that Alice and Charlie are these two dishonest participants. In **Step 6**, Charlie obtains $q_{Ab}$ and sends it to Alice. Subsequently, Alice performs Bell measurement on $\{p_A q_{Ab}\}$ to obtain $K_B \oplus R_B$. For example, if the measurement result of the $i$ th pair of $\{p_A q_{Ab}\}$ is $|\Phi^+\rangle$, Alice and Charlie can deduce that Bob's unitary operation is $I$, which means $K_B^i \oplus R_B^i = 0$. Otherwise, if the measurement result is $|\Psi^+\rangle$, they can deduce that $K_B^i \oplus R_B^i = 1$. At the same time, Alice sends $q''_{Bca}$ to Bob. So Alice and Charlie can obtain the value of $K_B \oplus R_B$ in **Step 6**, and Bob cannot detect it. After this, in **Step 8**, Bob announces the details of the additional unitary operation $R_B$. At this time, Alice and Charlie can compute $K_B = (K_B \oplus R_B) \oplus R_B$ to obtain Bob's private key $K_B$.

If Alice and Charlie want to manipulate the final secret key, they can control the values of $R_A$ or $R_C$ without being detected by Bob. After receiving the details of the permutation operation $\prod_A^{2n''}$ in **Step 8**, Bob rearranges the received sequence to obtain $q_{Bca}$ and performs Bell measurement on $\{p_B, q_{Bca}\}$ to obtain the measurement result $M_B$. Subsequently, Bob uses $K_{A\oplus C} = M_B \oplus R_A \oplus R_C$ to get the OR values of Alice and Charlie. However, if Bob gets a different additional unitary operation $R'_A$ or $R'_C$, then he will get a different values $K'_{A\oplus C}$. Hence, Bob will get a manipulated final secret key without detection.

As an example, we use the generation process of 2-bit key to explain the above attack. Suppose that $K_A=11$, $K_B=10$, $K_C=00$, $R_A=00$, $R_B=01$ and $R_C=11$. As above attack shows that Alice and Charlie can obtain $K_B \oplus R_B=11$ in **Step 6**. After receiving $R_B=01$ in **Step 8**, Alice and Charlie can get the value of $K_B$, $K_B=(K_B \oplus R_B) \oplus R_B=(10 \oplus 01) \oplus 01=10$. At this moment, Alice and Charlie obtain the final secret key $K=K_A \oplus K_B \oplus K_C = 11 \oplus 10 \oplus 00 = 01$. If they want transform the final secret key from $K=01$ to $K=11$, Alice announces $R_A=00$ and Charlie announces $R'_C=01$. After Bob obtain the value of $M_B$, $M_B=K_A \oplus K_C \oplus R_A \oplus R_C = 00$, he computes the final key as $K=M_B \oplus R_A \oplus R'_C \oplus K_B = 00 \oplus 00 \oplus 01 \oplus 10 = 11$. It denotes that Bob gets a manipulated final key without detection. The above analysis shows that Zhu et al.'s protocol cannot achieve the fairness property.

## 4 Conclusions

This article points out a Participant Colluding Attack on Zhu et al.'s quantum key agreement protocol, where two dishonest participants can manipulate the final secret key without being detected by the other participants. In this regard, Zhu et al.'s QKA protocol cannot achieve the fairness property.

## Acknowledgement

We would like to thank the Ministry of Science and Technology of the Republic of China, Taiwan for partially supporting this research in finance under the Contract No. MOST 104-2221-E-006-102 -.

# References

1. Bennett Ch H. and Brassard G. *Quantum cryptography: public key distribution and coin tossing Int*. in *Conf. on Computers, Systems and Signal Processing (Bangalore, India, Dec. 1984)*. 1984.
2. Hwang T., Lee K.-C., and Li C.-M., *Provably secure three-party authenticated quantum key distribution protocols.* Dependable and Secure Computing, IEEE Transactions on, 2007. **4**(1): p. 71-80.
3. Lo H.-K., Ma X., and Chen K., *Decoy state quantum key distribution.* Physical review letters, 2005. **94**(23): p. 230504.
4. Long G.-L. and Liu X.-S., *Theoretically efficient high-capacity quantum-key-distribution scheme.* Physical Review A, 2002. **65**(3): p. 032302.
5. Grosshans F., Van Assche G., Wenger J., Brouri R., Cerf N.J., and Grangier P., *Quantum key distribution using gaussian-modulated coherent states.* Nature, 2003. **421**(6920): p. 238-241.
6. Ouellette J., *Quantum key distribution.* Industrial Physicist, 2004. **10**(6): p. 22-25.
7. Renner R., *Security of quantum key distribution.* International Journal of Quantum Information, 2008. **6**(01): p. 1-127.
8. Lim C.C.W., Portmann C., Tomamichel M., Renner R., and Gisin N., *Device-independent quantum key distribution with local Bell test.* Physical Review X, 2013. **3**(3): p. 031006.
9. Lo H.-K., Curty M., and Qi B., *Measurement-device-independent quantum key distribution.* Physical review letters, 2012. **108**(13): p. 130503.
10. Zhou N., Zeng G., and Xiong J., *Quantum key agreement protocol.* Electronics Letters, 2004. **40**(18): p. 1149-1150.
11. Tsai C.-W., Chong S.-K., and Hwang T. *Comment on quantum key agreement protocol with maximally entangled states*. in *Proceedings of the 20th Cryptology and Information Security Conference (CISC 2010)*. 2010.
12. Chong S.-K. and Hwang T., *Quantum key agreement protocol based on BB84.* Optics Communications, 2010. **283**(6): p. 1192-1195.
13. Shi R.-H. and Zhong H., *Multi-party quantum key agreement with bell states and bell measurements.* Quantum information processing, 2013. **12**(2): p. 921-932.
14. Liu B., Gao F., Huang W., and Wen Q.-y., *Multiparty quantum key agreement with single particles.* Quantum information processing, 2013. **12**(4): p. 1797-1805.
15. Sun Z., Zhang C., Wang B., Li Q., and Long D., *Improvements on "Multiparty*

*quantum key agreement with single particles".* Quantum information processing, 2013. **12**(11): p. 3411-3420.
16. Huang W., Wen Q.-Y., Liu B., Su Q., and Gao F., *Cryptanalysis of a multi-party quantum key agreement protocol with single particles.* Quantum information processing, 2014. **13**(7): p. 1651-1657.
17. Shukla C., Alam N., and Pathak A., *Protocols of quantum key agreement solely using Bell states and Bell measurement.* Quantum Information Processing, 2014. **13**(11): p. 2391-2405.
18. Zhu Z.-C., Hu A.-Q., and Fu A.-M., *Improving the security of protocols of quantum key agreement solely using Bell states and Bell measurement.* Quantum Information Processing, 2015. **14**(11): p. 4245-4254.